# Turbulence in Zeeman Measurements from Molecular Clouds

Zhuo Cao and Hua-bai Li*
**Department of Physics, The Chinese University of Hong Kong, Shatin, Hong Kong**
*E-mail: hbli@cuhk.edu.hk

**Abstract**
Magnetic fields (B-fields) play an important role in molecular cloud fragmentation and star formation, but are very difficult to detect. The temporal correlation between the field strength ($B$) and gas density ($n$) of an isolated cloud has been suggested as an indication of the dynamical importance of B-fields relative to self-gravity. This temporal $B$-$n$ relation is, however, unobservable. What can be observed using Zeeman measurements are the "spatial $B$-$n$ relations" from the current plane of the sky. Nevertheless, the temporal $B$-$n$ relation argument has still been widely used to interpret observations. Here we present the first numerical test of the legitimacy of this interpretation. From a simulation that can reproduce the observed Zeeman spatial $B \propto n^{2/3}$ relation, we found that temporal $B$-$n$ relations of individual cores bear no resemblance to the spatial $B$-$n$ relations. This result inspired us to discover that the true mechanism behind the 2/3 index is random turbulence compression instead of symmetrical gravitational contraction.

### 1. Introduction

Star formation textbooks have often adduced the example of the collapse of an isolated cloud with uniform B-fields to argue that the $B$-$n$ relation should follow a power law (Fig. 1). In this model, $B$ is expected to be independent of $n$ in cases where B-field energy is absolutely dominant, as the Lorentz force will limit gas contraction along the field, so that gas cannot compress the field lines to increase $B$. In the opposite scenario, when self-gravity presides, the contraction will be isotropic and result in $B \propto n^{2/3}$, as only the contraction in the two dimensions perpendicular to the field lines can enhance $B$ (Fig. 1, left panel). In other words, the index of $n$ varies between 0 and 2/3 depending on the dynamical importance of B-fields (e.g., Li 2021; Crutcher et al. 2010).

However, it is impossible to monitor the temporal $B$-$n$ relation in reality. The published cloud $B$-$n$ relations depend on either sampling clouds with various mean densities (Crutcher et al. 2010; Jiang, Li & Fan 2020) or sampling various densities within a single cloud (e.g., Li et al. 2015; Kandori et al. 2018; Wang et al. 2020). In each case, $B$ and $n$ are surveyed from different spatial positions, and do not involve temporal developments from individual clouds. Moreover, a real cloud core is never really "isolated" and can exchange mass with its envelope due to, for example, turbulence.

We note with concern that in most extant literature, whether observational studies or simulations, the spatial $B$-$n$ relation indices have been interpreted using the temporal indexes as posited in the "textbook model". Here, we examine with ideal magnetohydrodynamic (MHD) simulations whether it is reasonable to interpret observations with temporal relations. In section 2, we review briefly the simulations, and in the Appendix, they are described in more detail. A summary of the results is presented in section 3. In section 4, we will examine the true reason behind the $B$-$n$ relationship of a magnetized turbulent cloud.

### 2. Simulations

The setup of the cloud simulation is largely adopted from (Zhang et al. 2019) and is detailed in the Appendix. Briefly, we simulated an isothermal cloud volume under periodic boundary conditions, starting with uniform density, uniform B-field, super-sonic/sub-Alfvenic turbulence driving, and slightly magnetically supercritical mass (Table I). The gravity was "turned on" after the turbulent energy was saturated. This resulted in dense cores of a few solar mass with $n$ peaking at ~$10^5$/cc (Fig. 2). Intriguingly, these cores turned trans- to super-Alfvenic (Table I) due to density-enhanced turbulent energy. This simulation accurately reproduced every major observation related to molecular cloud B-fields, including the "ordered cloud B-fields" (Li et al. 2015; Li et al. 2009; Planck Collaboration 2016), the significantly "deviated core B-fields" (Zhang et al. 2019; Zhang et al. 2014; Hull et al. 2014) (Table I), and, most importantly, the 2/3 index of the spatial $B$-$n$ relation (Crutcher et al. 2010; Jiang, Li & Fan 2020) (Fig. 2).

We aimed to elucidate the temporal $B$-$n$ relation of cloud cores. For this purpose, we adjusted the simulation setup in Zhang et al. (2019) by halting the turbulence driving after the energy was saturated and the gravity was turned on (see Appendix A1) (had we not done so, even bounded structures would have been dispersed shortly by the artificial driving, leaving no room for a "temporal" study). An interesting observation is that, although the overall kinetic energy became more sub-Alfvenic after turbulence driving was stopped, the dense cores remained trans- to super-Alfvenic (Table I) due to density enhancement.

Following the process summarized in Appendix A2, we found three "long-lived" cores (Fig. 1). Their temporal $B$-$n$ relations are plotted with the spatial relation in Fig. 2. Unlike the textbook model, here the cores undergo dynamic changes in both positions and shapes. We need to monitor the development closely to follow the position of the core peaks and use a fixed mass (two solar mass) to define the core volumes for measuring the mean $B$ and $n$. The textbook model (Fig. 1, left panel) also has a fixed mass; just here the fixed mass may not be composed of the same group of gas, as the gas can flow in and out of the boundary containing the fixed mass.

### 3. Results

The temporal relation not only varied from core to core, but also significantly differed from the spatial relation. Therefore, the 2/3 index of the spatial relation certainly needs another interpretation than that provided in the textbook temporal model. To see this from another perspective, we note that the spatial $B$-$n$ relation before the gravity was turned on (red in Fig. 2) already possessed an index of 2/3 for $n > 10^4$/cc. This 2/3 index is due to "random" compression of B-field lines by *super-Alfvenic turbulence* inside of cloud cores instead of isotropic compression by self-gravity as in the textbook model. We emphasize that the turbulence energy driven at large scale is sub-Alfvenic, and the 2/3 index only occurs at core densities ($n > 10^4$/cc; Fig. 2), where turbulence energy is amplified by the density enhancement to slightly super-Alfvenic (Table I) (Li 2021; Zhang et al. 2019).

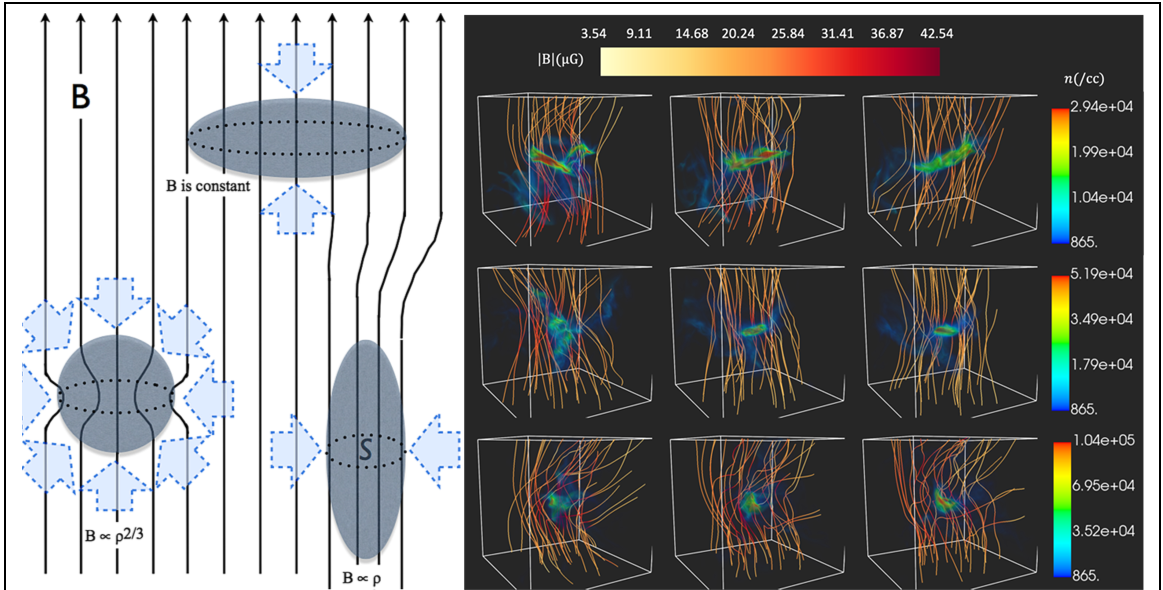

**Fig. 1. A comparison between the cloud cores from the idealized "textbook" model and from MHD simulations.** *Left*: The textbook model of cloud core $B$-$n$ relations (adopted from Li 2021). The arrows indicate the direction of effective compression. The case with dominant B-field or self-gravity will possess an index of 0 or 2/3, respectively. An index larger than 2/3 can result from non-gravitational forces, e.g., turbulence, stellar wind, or supernova compression. *Right:* Cores from MHD simulations appear markedly different from the textbook model. Each row presents the data from one core; from top to bottom are Core 1 to 3. Each column is derived from one snapshot of each core; from left to right are starting, middle, and ending snapshots, which are defined in Appendix A2.

**Table I**

|  | Core 1 | Core 2 | Core 3 | Cloud, end of turbulence driving | Cloud, after core formation |
|---|---|---|---|---|---|
| $M$, Mach number | 4.4 | 1.1 | 2.4 | 5.6 | 3.8 |
| $M_A$, Alfvén Mach number | 3.1 | 0.9 | 1.8 | 0.60 | 0.44 |
| Magnetic criticality | 1.5 | 2.2 | 2.0 | 2.0 | 2.0 |
| B deviation from the initial condition | 10° | 12° | 48° | 0° | 0° |

**Note:** $M = \langle v/c_s \rangle$; $M_A = \langle v/v_{Alf} \rangle$. The last row indicates the angle between the mean magnetic field and the initial magnetic field (along the z-axis). The last two columns are from the entire simulation domain, which are sub-Alfvénic, while the cores are trans- to super-Alfvénic.

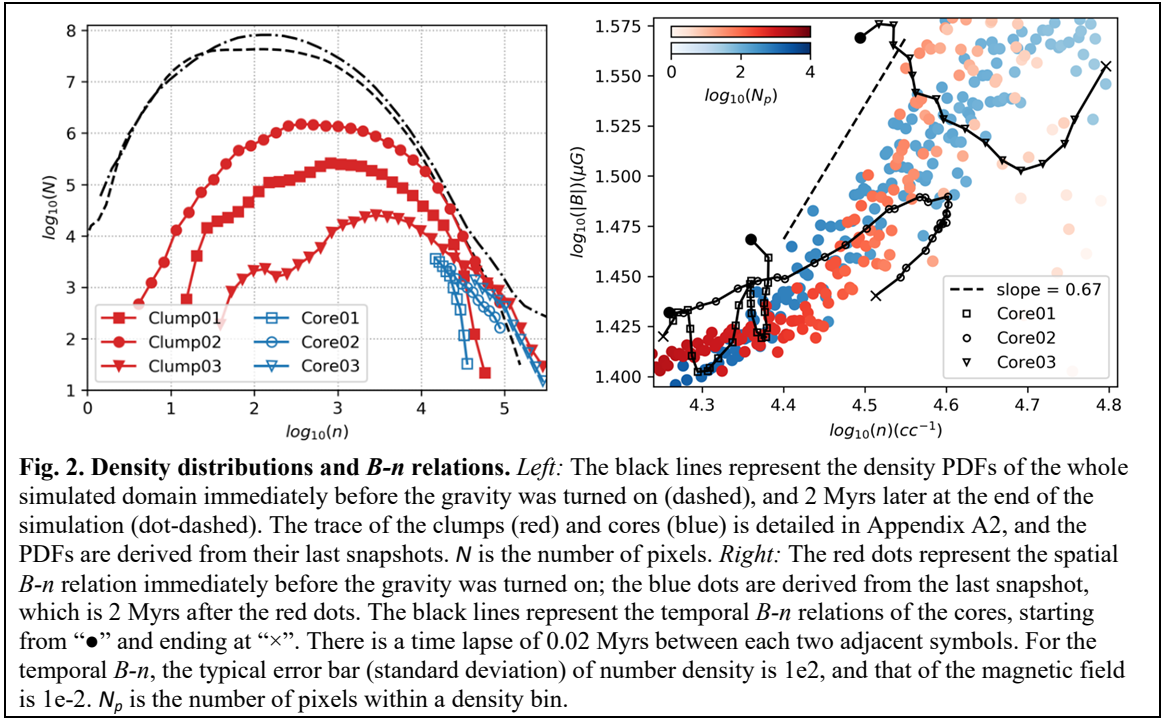

**Fig. 2. Density distributions and *B-n* relations.** *Left:* The black lines represent the density PDFs of the whole simulated domain immediately before the gravity was turned on (dashed), and 2 Myrs later at the end of the simulation (dot-dashed). The trace of the clumps (red) and cores (blue) is detailed in Appendix A2, and the PDFs are derived from their last snapshots. *N* is the number of pixels. *Right:* The red dots represent the spatial *B-n* relation immediately before the gravity was turned on; the blue dots are derived from the last snapshot, which is 2 Myrs after the red dots. The black lines represent the temporal *B-n* relations of the cores, starting from "●" and ending at "×". There is a time lapse of 0.02 Myrs between each two adjacent symbols. For the temporal *B-n*, the typical error bar (standard deviation) of number density is 1e2, and that of the magnetic field is 1e-2. $N_p$ is the number of pixels within a density bin.

Furthermore, another type of spatial *B-n* relation based on the *B* and *n* profiles within individual clouds/cores has also been observed, and a wide range of indices between 2/5 and 2/3 has been reported (e.g., Li et al. 2015; Kandori et al. 2018; Wang et al. 2020). In Fig. 3, this type of *B-n* relation is shown for the three cores in our simulation. It appears that the indices are closer to the lower end of the observed range. It is worth noting, however, that Fig. 3 is produced by binning the simulation pixels by *n* and obtaining the mean *B* from each bin. Observers, on the other hand, must "somehow" estimate the mean *n* from each line of sight, which usually covers a broad range of *n*. Moreover, observers usually use the DCF method when estimating *B* (e.g., Chen et al. 2022; Liu, Qiu & Zhang 2022), which is also not taken into account when plotting Fig. 3. It is beyond the scope of this article to make a detailed comparison between Fig. 3 and the observations. The point that we wish to make is that the *B-n* spatial relations obtained from individual cores (Fig. 3) do not reflect the *B-n* temporal relations (Fig. 2, right panel) even without the observational "lens", nor should they be interpreted using the textbook temporal model.

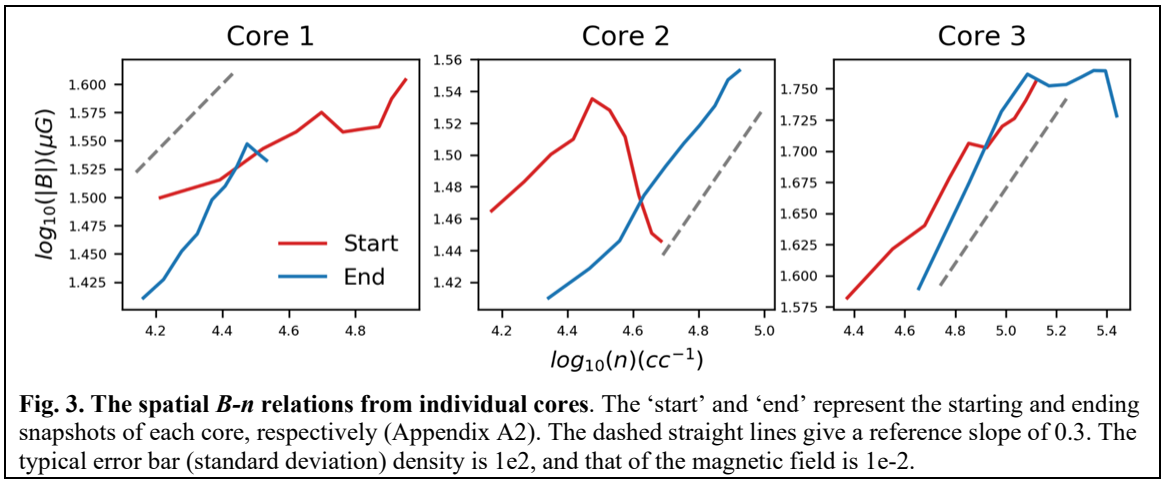

**Fig. 3. The spatial *B-n* relations from individual cores.** The 'start' and 'end' represent the starting and ending snapshots of each core, respectively (Appendix A2). The dashed straight lines give a reference slope of 0.3. The typical error bar (standard deviation) density is 1e2, and that of the magnetic field is 1e-2.

## 4. Discussion

In the following, we take a closer look at the three cores to better understand their temporal $B$-$n$ relations. They are all among the highest densities in the simulation, as illustrated by the density distributions in Fig. 2. For Core 1, the density maps in Fig. 1 show a trend of decreasing density (the red-color region decreases as time passes), which is also indicated by the temporal plots (Figs. 2 and 3). Conversely, the density of Core 3 increased with time, and the $B$-$n$ relation index was negative at the beginning but became 2/3 at the end. Core 2 started with a shallow $B$-$n$ relation and the $B$ and $n$ "bounced" back in the later stage, which appears as a turning point in the $B$-$n$ plot. The presence of turbulence makes the expectation of a constant $B$-$n$ index (Fig. 1) unrealistic.

### 4.1 Force interactions

The "textbook" argument for the $B$-$n$ relation indices solely depends on gravity to concentrate gas, and the B-field can only play a role against gravity (Fig. 1). In reality, however, thermal pressure should help support the cloud, and turbulence can either disperse or enhance the densities of gas and/or field-lines. As a consequence, even the temporal $B$-$n$ relation should not be interpreted by the textbook model before a careful examination of the model assumptions. Next, we check whether the gas flows, which shape the $B$-$n$ relation, are converged only by self-gravity and hindered by B-field forces.

Only forces perpendicular to the B-field, $f\perp$, can regulate the field strength. Positive/negative divergence of $f\perp$ can accelerate the gas to disperse/compress field lines, thereby decreasing/increasing the field strength. To determine what force is mainly compressing or dispersing core fields, we can integrate the divergence of $f\perp$ within the core; the results are shown in Fig. 4, where, for simplicity, $f\perp$ is defined by the mean-field direction of a core.

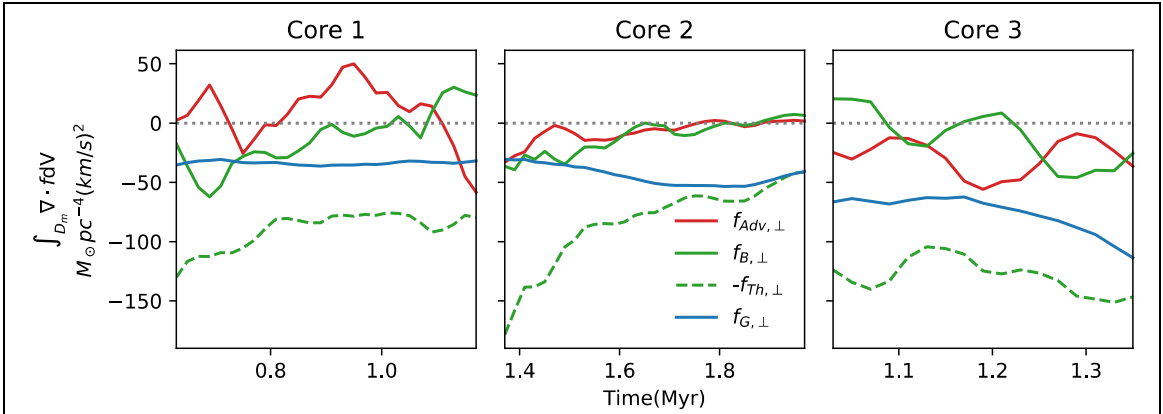

**Fig. 4. How the forces regulate B-field strength.** This figure illustrates the volume integration of the divergences of the force $f_i\perp$ in the volume above the mean density of a core (details of the calculation can be found in *Methods*), where "$i$" can be $G$, $Th$, $B$, or $Adv$, corresponding to gravitational, thermal, magnetic, and advection, respectively. "$\perp$" indicates that only the component perpendicular to the mean field is considered here, as only the perpendicular component is capable of affecting the strength of the field. Note that the negative of $f_{Th}\perp$ divergence is plotted to conserve space, as most of the divergences of other forces are negative. A positive/negative $f_i\perp$ means that, in an average sense, the force was dispersing/compressing the field lines.

As expected, the gravitational force always compressed the field (with negative $f_G\perp$ divergence in Fig. 4); whereas, the thermal pressure constantly opposed gravity with a positive divergence (note that plotted in Fig. 4 is negative $f_{Th}\perp$ divergences to save space). What should be noticed is that, most of the time, the magnitude of $f_{Th}\perp$ divergence is significantly greater than that of $f_G\perp$. In other words, it is gravity that needs assistance against thermal pressure in order to hold a core together. This differs greatly from the textbook model (Fig. 1), in which the thermal force is assumed to be negligible. Indeed, the divergence of the advection and B-field forces ($f_{Adv}\perp$ and $f_B\perp$ in Fig. 4, respectively) are negative for significant periods of time, when they confined the cores rather than supported the cores against gravity. It is worth noting that the simulation is isothermal, and the turbulence remains supersonic in the cores (Table I). It is the density gradients set up by turbulent shocks that are responsible for the $f_{Th}\perp$. This is in agreement with the scenario that the 2/3 index is also due to the random turbulence compression. Only the densest Core 3 survived until violating the Truelove criterion (Truelove et al. 1997), while the other two cores are dynamic transient structures. B-fields play more of a stabilizing role in all cases. In Fig. 4, the divergences of -$f_{Th}\perp$ and $f_B\perp$ (the two green lines) share the same trend, i.e., when $f_{Th}\perp$ became more dispersive, $f_B\perp$ turned more compressive, and vice versa. In addition, B-fields also reacted against the advection force, which can be observed from the fact that the higher-frequency variations of the divergences of $f_{Adv}\perp$ and $f_B\perp$ are usually complementary. This is especially apparent for Core 3.

### 4.2 Virial parameters

In assessing the gravitational boundness of molecular clouds, observers estimate the Virial parameter (Bertoldi & McKee 1992), $\alpha = 2E_{Tu}/E_G$ (two times the turbulence to gravitational energy ratio; see the Appendix A2 for detailed definitions of $E_{Tu}$ and $E_G$), which usually has a negative power law relationship with the core mass (e.g., Keown et al. 2017; Kirk et al. 2017; Kerr et al. 2019). Cores 1-3 have $\alpha$'s close to one, which are consistent with observations (Figure 5). Note that $\alpha$ ignores thermal and magnetic energy, so $\alpha < 1$ does not guarantee a contraction. Our simulation also replicated other observations, such as magnetic criticality = 1-2 (Table I; Li et al. 2013; Li et al. 2015; Myers & Basu 2021) and core density profiles $n \propto r^{-1.46\pm0.12}$ (Zhang et al. 2019, Pirogov 2009; Kurono et al. 2013) leading to strong pressure gradients (Fig. 4). With the marginal Virial parameter, magnetic criticality, and strong density gradient, cores are on the verge of contraction and expansion, as demonstrated by Cores 1 (expanding), 2 (bouncing), and 3 (contracting). This is another perspective to avoid using the ever-collapsing textbook model (Fig. 1, left) to interpret observations.

### 4.3 Magnetic field diffusion

Finally, we note that turbulence may induce both ambipolar (Li & Houde 2008; Tang, Li & Lee 2018) and/or reconnection (Lazarian et al. 2012) diffusion, which are not simulated here but can potentially modify $B$-$n$ relations. This study, however, is not intended to predict the exact index of $B$-$n$ relations, but rather aims to explore the rationale behind accessing temporal $B$-$n$ relations via Zeeman measurements (spatial relations). Field diffusion should only flatten both the temporal and spatial $B$-$n$ relations to some extent (see, e.g., Tsukamoto et al. 2015) instead of resolving their significant discrepancy (Figure 2).

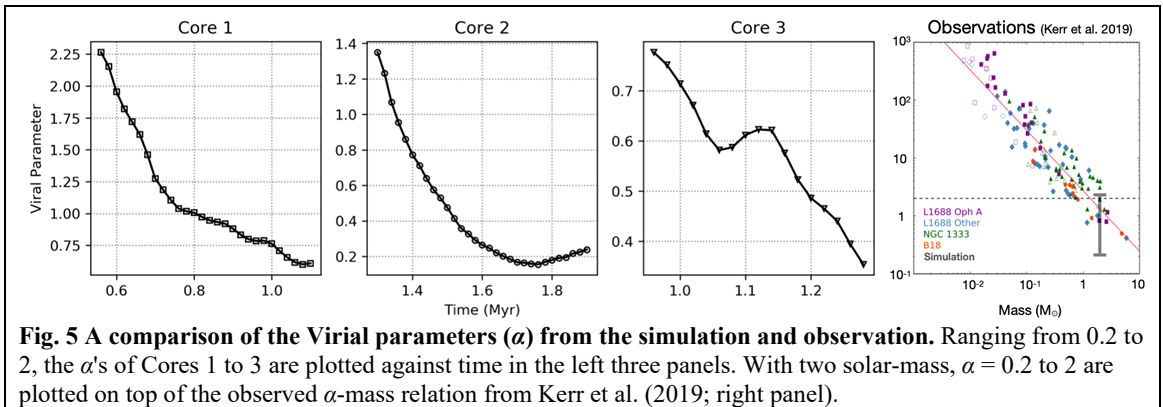

**Fig. 5 A comparison of the Virial parameters ($\alpha$) from the simulation and observation.** Ranging from 0.2 to 2, the $\alpha$'s of Cores 1 to 3 are plotted against time in the left three panels. With two solar-mass, $\alpha$ = 0.2 to 2 are plotted on top of the observed $\alpha$-mass relation from Kerr et al. (2019; right panel).

### 5. Conclusion

Since magnetized turbulent clouds exhibit significant discrepancies in their spatial and temporal $B$-$n$ relations, the 2/3 index inferred from Zeeman measurements (Crutcher et al. 2010; Jiang, Li & Fan 2020) should not be interpreted as highly magnetic supercritical. The puzzle (Pattle et al. 2023) of the coexistence of a 2/3 index and ordered B-fields, which is a sign of magnetic sub- to trans-criticality, has thus now been solved. In fact, recent observations of *magnetic field-column density* relations suggest that cloud cores are only slightly super-critical (Li et al. 2013; Myers & Basu 2021), which is also in accordance with the simulation presented here (Table I). As indicated in Table I, within cloud cores ($n > 10^4$/cc), turbulence can be super-Alfvenic to randomly compress B-fields, and thus result in the 2/3 index. We learned from "Dynamic Cores in Hydrostatic Disguise" (Ballesteros-Paredes, Klessen & Vázquez-Semadeni 2003) two decades ago that a Bonnor-Ebert-like density profile does not necessarily indicate structural stability. Our study suggests that, in the past decade, turbulence likely misled us again by leaving a "2/3" footprint in the $B$-$n$ relation.

# Appendix

### A1. Simulation Setup

We simulated the interior of a molecular cloud using the ideal magnetohydrodynamic (MHD) code ZEUS-MP (Hayes et al. 2006; Otto, Ji & Li 2017). Assuming isothermal, the code solves the following set of equations:

$$\frac{\partial \rho}{\partial t} + \nabla \cdot (\rho \boldsymbol{v}) = 0$$

$$\rho \left(\frac{\partial}{\partial t} + \boldsymbol{v} \cdot \nabla\right) \boldsymbol{v} = \boldsymbol{J} \times \boldsymbol{B} - \nabla p - \nabla \phi$$

$$\frac{\partial \boldsymbol{B}}{\partial t} = \nabla \times (\boldsymbol{v} \times \boldsymbol{B})$$

$$p = \rho c_s^2$$

$$\boldsymbol{J} = \nabla \times \boldsymbol{B}$$

$$\nabla \cdot \boldsymbol{B} = 0$$

$$\nabla^2 \phi = 4\pi G \rho$$

where $\rho$ and $p$ are mass density and thermal pressure, respectively; $\boldsymbol{v}$ and $\boldsymbol{B}$ are velocity and magnetic field vector, respectively; the constant $c_s \sim 0.2$ km/s is the sound speed assuming a temperature of 10 K; and $G = 4.3 \times 10^{-3} \text{ pc} \cdot M_\odot^{-1} \cdot \left(\frac{km}{s}\right)^2$ is the gravitational constant.

The initial B-field is 14.4 $\mu$G in the $\boldsymbol{z}$-direction uniformly. The ratio between thermal pressure and magnetic pressure is $\beta = 0.05$. The uniform initial density is $\rho = 1.2 \times 10^{21}$ g cm$^{-3}$ or $n = 300$ H$_2$/cc, assuming a mean molecular weight of 2.36. The size of the simulation domain is $4.8^3$ pc$^3$, resolved by $960^3$ cells, and a periodic boundary condition is applied. Correspondingly, the mass of the whole domain is twice the magnetic critical mass ($M_\Phi = \Phi/2\pi G^{1/2}$, where $\Phi$ is magnetic flux).

The simulation was separated into two phases. During the first phase, turbulence was driven without self-gravity. When the turbulent energy was saturated (i.e., when the turbulent energy spectrum is stable), the Mach number of turbulence ($M$) was 5.6 (Table 1) and the rms Alfven Mach number ($M_A$) was 0.6. Self-gravity was turned on in the second phase, while the turbulence driving was halted. The simulation was terminated after gravity had been turned on for 2 Myrs, before any core can violate the Truelove criterion (Truelove et al. 1997).

A detailed description of turbulence driving is given in (Otto, Ji & Li 2017). Briefly, we set up a vector field $\boldsymbol{a}(\boldsymbol{k})$ in the Fourier space with a zero mean and a variance $\propto k^6 exp(8k/k_c)$. Therefore, the power spectrum peaks sharply at $k_c$, which we set as 2, corresponding to a driving scale of 2.4 pc. We then apply inverse Fourier transform on vector field $\boldsymbol{a}(\boldsymbol{k})$ with a random phase to obtain the driving velocity field $\boldsymbol{v}(\boldsymbol{x})$. Details of the saturated velocity spectrum are given in (Zhang et al. 2019) (see their Fig. 3). In this simulation, the turbulence driving is purely solenoidal.

### A2. Core Finder

As stated in the main text, our goal is to study the temporal *B-n* relation of individual molecular cloud cores. This requires the data of a core at different evolution stages. Defining a core in a single simulation snapshot is simple. The difficulty lies in tracing this core in other snapshots due to transposition and deformation. Here, we describe how to trace a core across snapshots.

Prior to core extraction, we first define "clumps", as dense clumps are the birth beds of cores. We define clumps using potential field contour surfaces, such that, within a clump C, $\frac{|E_G|}{E_{Tu}+E_B+E_{Th}} > 1/2$, which is a relaxed gravitational bounded condition. In the equation, $E_{Tu} = \frac{1}{2}M\sigma^2$ indicates the turbulent energy. The energies $E_B = \frac{1}{8\pi}\int_V B^2 dV$ and $E_{Th} = \frac{3}{2}Mc_s^2$ are magnetic field and thermal energy, respectively. Finally, the

$E_G = \frac{1}{2}\int_V \rho\phi dV$ is the potential energy, where the potential field $\phi$ is due to all of the mass under the periodic boundary condition. We assume the global maximum of potential as the zero potential point.

In accordance with the above criterion, we scanned through potential contour surfaces in each snapshot to look for clumps. From high to low, we scanned through 40 potential values evenly distributed between zero and the minimum potential. For each potential value, the connected component labelling (CCL) method (Silversmith 2021) was applied to index individual volumes confined by the contour surfaces. The clump criterion was applied on each of these volumes, and clumps were excluded from further examination with lower potential values. Two clumps in adjacent snapshots are considered to be temporally connected if they spatially overlap. The algorithm is summarized in Fig. 6 below.

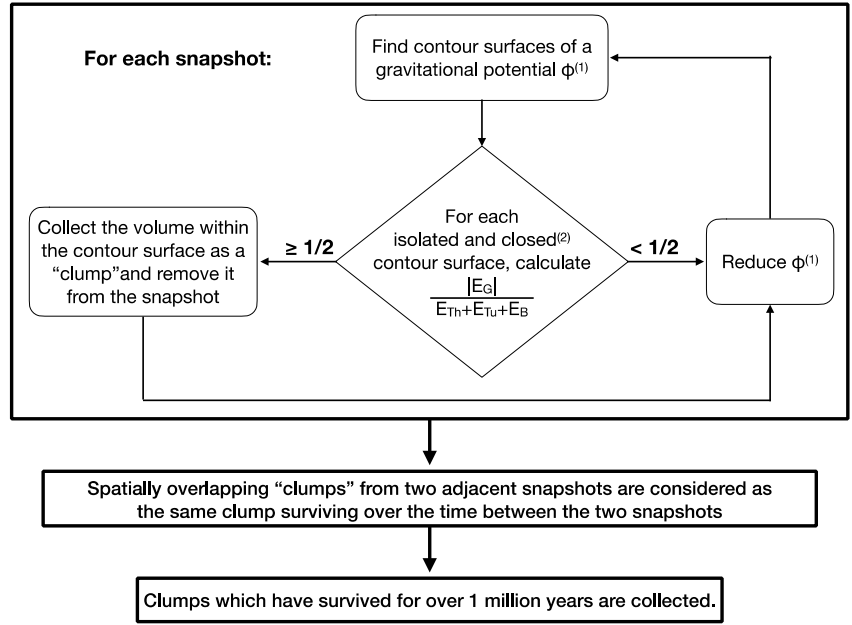

**Fig. 6 The clump-finding algorithm.** (1) From zero to the minimum potential, 1/40 of this potential range is reduced in each step. (2) Defined by the CCL method.

Several sets of temporally connected clumps are found. Only clumps persisting for more than 1 Myrs are retained for further analysis, because only long-lived clumps can host long-lived cores for temporal study. Three clumps meet the requirements.

We define a core with fixed mass as the following. The mass of a core is typically 0.1 to 10 solar masses (Mac Low & Klessen 2004); we use two solar masses in this study. For each clump, the snapshot with the highest peak density is identified, and the core is defined by the density contour surface containing this peak and two solar masses. The algorithm to identify the same core in the earlier and later snapshots is detailed in Fig. 7. Although a core can transpose the position and deform the shape, the difference in adjacent snapshots is expected to be small as long as the time difference is relatively short. For this purpose, we took frequent snapshots every 0.02 Myrs. The algorithm in Fig. 7 requires the center of mass (com) of a core to always fall within the effective core scale from the projected com position defined by the position and velocity of the com in the previous snapshot; the effective core scale is defined by two times of $r = \left(\frac{3}{4\pi}V\right)^{1/3}$, where $V$ is the core volume in the previous snapshot.

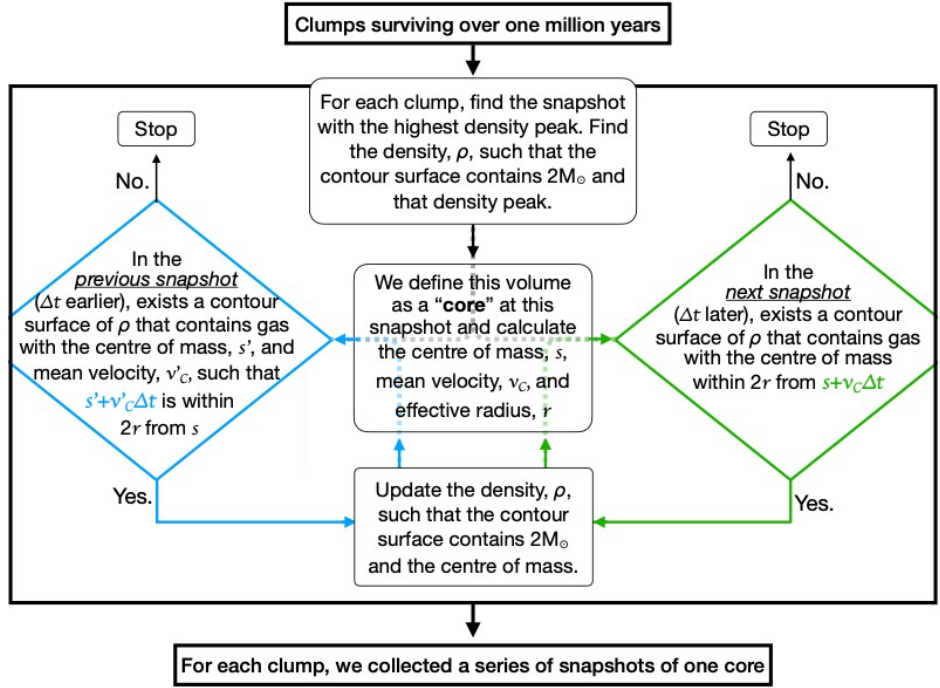

**Fig. 7 The core-finding algorithm.**

### A3. Integrated Divergence

A major focus of our analysis is the evolution of the B-field strength, which is driven by force components perpendicular to the local mean field $\bar{B}$:

$$\boldsymbol{f}_{i,\perp} = \boldsymbol{f}_i - \boldsymbol{f}_{i,\parallel} = \boldsymbol{f}_i - \boldsymbol{f}_i \cdot \frac{\bar{\boldsymbol{B}}}{|\bar{\boldsymbol{B}}|}$$

, where $i = B, Adv, Th$, or $G$, standing for the *Lorentz*, *advection*, *thermal*, and *gravitational* force, respectively. More precisely, they are formulated as follows:

$$\boldsymbol{f}_{Adv} = -\rho\,(\boldsymbol{v}\cdot\nabla)\,\boldsymbol{v}$$

$$\boldsymbol{f}_B = \boldsymbol{J}\times\boldsymbol{B} = (\nabla\times\boldsymbol{B})\times\boldsymbol{B}$$

$$\boldsymbol{f}_{Th} = -\nabla P$$

$$\boldsymbol{f}_G = -\rho\nabla\phi$$

To determine whether, in general, the field lines were compressed or dispersed by $\boldsymbol{f}_{i,\perp}$, we integrated the divergence within $D_m$, the core volume above the mean density:

$$\int_{D_m} \nabla\cdot\boldsymbol{f}_{i,\perp}\,dV$$

# Acknowledgement

This research is supported by General Research Fund grants from the Research Grants Council of Hong Kong: Nos. 14305717 and 14304616, and by Collaborative Research Fund grant No. C4012-20E.